# 10 Inventions on keyboard attachments
## -A TRIZ based analysis

**Umakant Mishra**

Bangalore, India
umakant@trizsite.tk
http://umakant.trizsite.tk

**Contents**

# 1. Introduction

A keyboard is the most important input device for a computer. It is quite essential for almost all types of computers ranging from servers, desktops, and laptops to small handheld computers like palmtops and PDAs. Generally larger computers have separate keyboards while portable computers have in-built keyboards. The keyboard consists of a set of keys, a key pressing mechanism and a connection to the computer.

Although the primary objective of the keyboard to input data into the computer, the advanced keyboards keep various other things in mind, such as, how to use the same keyboard for various other purposes, or how to use the same keyboard efficiently by using various other attachments to the keyboard. This objective led to various inventions on keyboard attachments, some of which are illustrated below in this article.

## 2. Inventions on keyboard attachment

### 2.1 Joystick attachment for computer keyboard (Patent 4575591)

**Background problem**

Computer games are played with keyboard, mouse, joystick or other input devices. A joystick is a specialized input device for the purpose and often more suitable and preferred for gaming than other devices.

Joystick is typically large to be comfortably held and operated by the player. This objective of a joystick does not allow it to be smaller. But practically it occupies lot of desk space. A computer table is hardly left with any space when a monitor, keyboard and joystick are in place.

**Solution provided by the invention**

Thomas Lugaresi disclosed a method of (Patent 4575591, Issued in March 86) attaching a joystick with a keyboard. The objective is to have a new and novel joystick attachment to the keyboard which can be attached and detached as required.

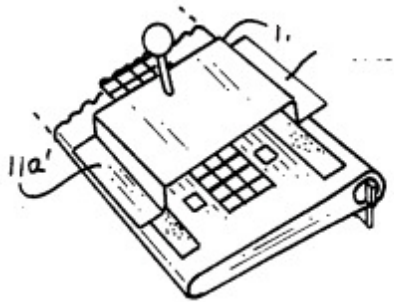

A joystick attachment for a computer keyboard which includes a bracket on which is supported a housing for a joystick assembly, the bracket being detachably mounted on a computer keyboard. The joystick is placed above the keyboard but with some gap from the upper surface of the keys for safety. The attachable and detachable feature gives the dual benefit of a keyboard and joystick without loosing extra space.

**TRIZ based solution**

The same keyboard is changed to a joystick when the user wants to play a game and changed to a keyboard when the wants to input characters **(Ideal Final Result)**.

The invention uses a joystick attachment to the keyboard, which is mounted on top of the keyboard without using additional table space **(Principle-7: Nesting).**

## 2.2 Computer key cover apparatus (Patent 5096317)

### Background problem
The computer keyboard is exposed to moisture, dust, smoke and external pollutions. When the keyboards are lying in disturbing environments like training units, it is necessary to prevent inadvertent depressing of keys of the keyboard. It is therefore required to have an improved cover for the keyboard.

### Solution provided by the patent
Phillippe invented a keyboard cover (patent 5096317, assigned to self, issued March 1992) where the cover apparatus includes a plurality of forward and rear side walls defining rows of openings, wherein each row of openings is arranged overlying rows of computer keys. The openings are provided with cover housings to selectively cover individual or the plurality of keys, wherein each of the plurality of keys of the computer keyboard is positioned within the openings to provide selective covering thereof to minimize inadvertent pressing of such keys in usage of the keyboard.

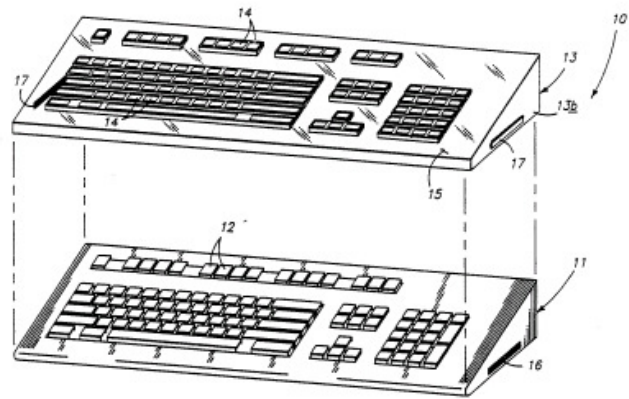

The invented cover is easy to manufacture and can be sold at low prices.

### TRIZ based analysis
The invention intends to provide a protective cover for the keyboard **(Principle-9: Prior Counteraction)**.

The cover takes a shape to protect selective keys of the keyboard from being inadvertently depressed **(Principle-3: Local Quality)**.

## 2.3 Keyboard with attachable pointing device (Patent 5281958)

### Background problem
A mouse takes quite a lot of space on the desktop and does not move with the movement of keyboard. When you pull the keyboard to your lap, there is no place for keeping the mouse. A laptop computer is often used in environments where a work surface is not provided. There is a need to move the pointing device according to the movement of the keyboard.

**Solution provided by the invention**

Ashmun et al. invented a keyboard (Patent 5281958, assigned to Microsoft, Issued in Jan 94) to which a pointing device is attachable through a clamp. The pointing device is attachable not only to the keyboard but even to any part of the computer such as the screen. The pointing device is coupled to an attachment or clamping assembly. The variable clamping assembly makes it suitable to fit into any type and size of different keyboards.

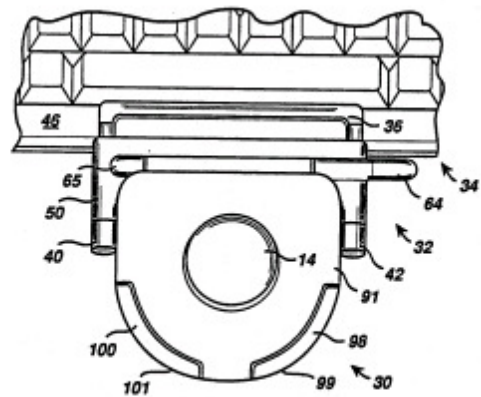

Figure 2

The full weight of the pointing device is supported by the clamps and the holding keyboard. This makes it movable with the keyboard and eliminates the need of extra working surface for the pointing device.

**TRIZ based analysis**

The pointing device should be placeable at any position of user comfort **(desired result)**.

The invention attaches the pointing device with the keyboard (or monitor or elsewhere) by using a clamp **(Principle-5: Merging)**.

This makes the position (or placement) of the pointing device flexible to be mounted at a convenient position instead of being placed only the table **(Principle-15: Dynamize, Principle-17: Another Dimension)**.

The pointing device can be placed at different sides of the keyboard and even on the monitor by using a clamp **(Principle-24: Intermediary, Principle-15: Dynamize)**.

**2.4 Clipboard attachments to computer keyboard (Patent 5786861)**

**Background problem**

Clipboards are used to hold paper for viewing while typing on a computer keyboard. Likewise, the keyboards also need dust covers. Cannot we combine the both and find out a way of using the clipboards as dust covers?

**Solution provided by the invention**

Yes, Joan Parker disclosed a method (US patent 5786861, Assigned to Joan Parker, Issued in July 98) to combine two clipboards and a dustcover into single unit which is attached to the keyboard to provide an easy view and easy reach using little disk space.

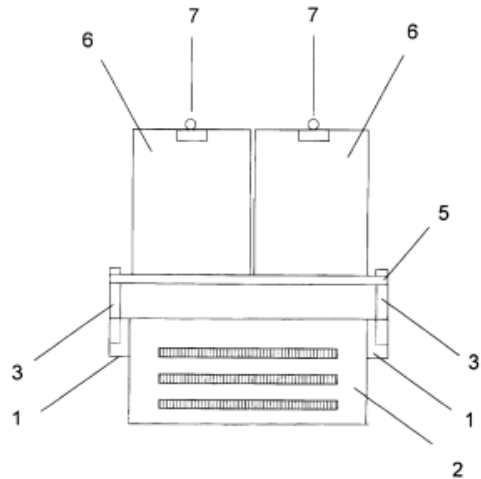

The invention includes two clipboards which give the convenience of holding two papers and enough cover plate for covering the keyboard. The clipboards are detachable so that the same clipboards can also be used as dustcover for the computer keyboard. The clipboards are also adjustable to hold papers in different viewing positions.

**TRIZ based solution**

Papers should be viewable while working with a computer **(desired result)**.

Keyboards should not accumulate dust **(desired result)**.

The invention uses clipboards, which acts for dual purposes, viz., for holding papers and for covering the keyboards **(Principle-6: Universality)**.

**2.5 Keyboard audio controls for integrated CD-ROM players (Patent 5881318)**

**Background problem**

The conventional CD ROMs do not provide adequate buttons to control the functions such as play, stop, next track, fast forward etc. The manufacturers of CD ROM drives expect the users to control the playing through the computer.

This method has a drawback that the program is not immediately available when the system is turned on. The program has to be loaded every time to operate the CD drive. There is a need to control the CD ROM player functions through some means without depending on the player software.

**Solution provided by the invention**

Liebenow disclosed a method (patent 5881318, assigned to Gateway 2000, Inc., March 99) to control the CD ROM through the keyboard. There will be a predefined sequence of keystrokes to control the CD ROM functions. When the predefined key combinations are pressed the OS receives the command from the BIOS to do the specific play or pause operation. This mechanism will not only make it independent of a player application. But also make the key combinations common even for different Operating System environments

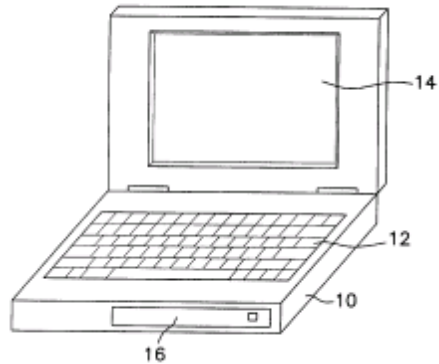

**TRIZ based analysis**

All controls to the CD ROM should be available without waiting for loading of any CD ROM software **(desired result)**.

The invention preloads the program in the BIOS and uses control keys on the keyboard to operate the CD ROM **(Principle-10: Prior Action)**.

This method facilitates using CD ROM through the same control keys irrespective of any change in the OS environment **(Principle-6: Universality)**.

**2.6 Keyboard having a storage device (patent 5892502)**

**Background problem**

Conventionally a keyboard is doing the job of a keyboard. When the keyboard is having so much of space inside it, the space is not utilized for any other purpose. Won't it be nice to use the available space inside keyboard for some useful purpose?

**Solution provided by the invention**

Hiller disclosed a method of integrating an optical disk drive under the numeric keypad area of the keyboard (US patent 5892502, April 99). The keyboard would have a headpone jack for the user to listen the music CD. This mechanism would make more effective use of the numeric keypad area.

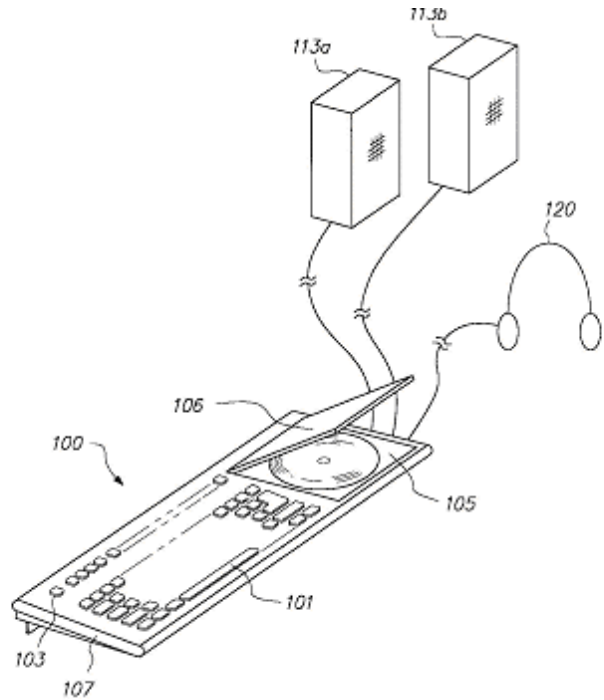

The invention is very useful as the user can change the CDs in a remote wireless keyboard without physically going to the CPU box which normally contains the CD drive.

**TRIZ based analysis**

The invention utilizes the available space inside the keyboard by putting the CD drive into it. As keyboard remains closer to the user, the user can easily replace the CDs without moving to the computer box. **(Principle-5: Merging, Principle-25: Self service)**.

**2.7 Multimedia console keyboard (patent 5892503)**

**Background problem**

The conventional keyboard does only the function of a keyboard. A multimedia computer needs a lot of special controls. Multimedia computers provide audio, video and computing. The control of audio and video functions, such as volume, brightness and contrast etc are done through separate keys attached with the device or through different software. This conventional arrangement is not very convenient for multimedia applications.

**Solution provided by the invention**

Seung-Min Kim disclosed a multimedia console keyboard (US Patent 5892503, assigned to AST Research, April 99) which provides multimedia functions easily accessible to the user. As per the invention the keyboard will provide speakers, microphones, volume control, brightness control, contrast control, a pointing device, and conventional PS/2 connectors on each side of the keyboard for left-handed or right-handed mouse connection.

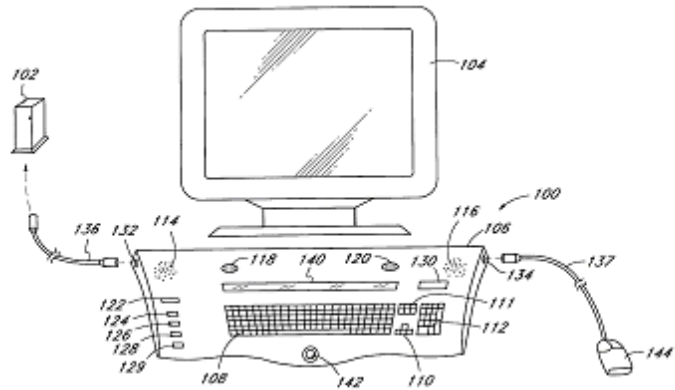

### TRIZ based analysis

The keyboard should have all features to support a multimedia computer **(desired result)**.

The invention includes necessary controls of the multimedia computer inside the keyboard **(Principle-38: Enrich)**.

### 2.8 Keyboard-compatible optical determination of object's position (Patent 5909210)

#### Background problem

There are several pointing devices including mouse, trackball, touchpad, touch screen, joystick and so on. Each has its own advantages and disadvantages. A mouse although convenient to hold, requires significant amount of space. The cable of a cabled mouse obstructs the movement of mouse. All pointing devices also need to move the hand from the keyboard to the pointing device which looses the flow of typing.

#### Solution provided by the invention

Knox et al, invented a method (patent 5909210, assigned to Compaq Computer Corporation, June 99) which communicates positional information to the computer by obstructing a light grid with the user's finger. The method provides an optical digitizer comprising a light grid. The light source and light sensor detects the obstacle made by user's finger. This method does not require the user to move its hands out of the keyboard, so enables the user's time to be more effectively utilized.

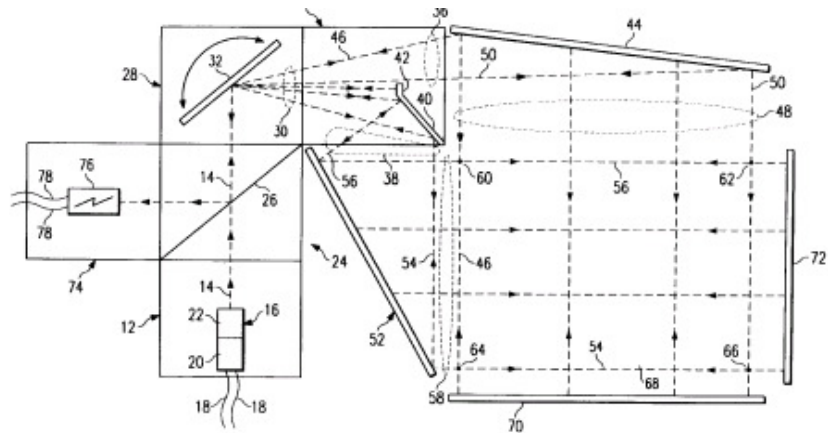

**TRIZ based analysis**

The user need not move hands from the keyboard to operate the pointing device **(Ideal Final Result)**.

The invention integrates a pointing device very much inside the keyboard like a control stick in other inventions **(Principle-5: Merging)**.

The method substitutes a manual control (of mouse or trackball) with an optical control **(Principle-28: Mechanics Substitution)**.

**2.9 Piano-style keyboard attachment for computer keyboard (Patent 5971635)**

**Background problem**

A PC is capable of producing sound of different musical organs. Although the PC can be used to play a musical organ, the keyboard of the PC is different form that of a musical organ like piano. There have been several attempts to assign some chords to different alphanumeric keys on a computer keyboard to play the music. But playing music by using alphanumeric keys neither gives a feeling of playing a musical organ nor gives the efficiency of a piano keyboard. How to solve this problem?

**Solution provided by the invention**

Herbert Wise disclosed a method of attaching a piano style keyboard to the computer keyboard. (US Patent Number 5971635, Assigned to Music Sales Corporation, Oct 99). According to the invention, the piano keyboard will be placed on top of the standard keyboard and fixed with anchors. Each piano-styled key has a protrusion extending downwardly for striking a corresponding computer key. The anchors comprise walls which extend deep in between computer keys in order to secure the apparatus to the computer keyboard.

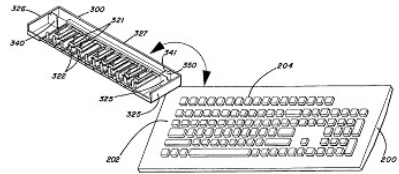 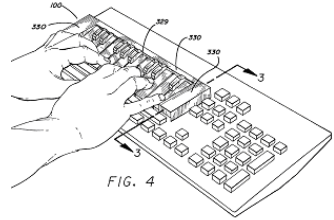

FIG. 4

**TRIZ based solution**

The computer keyboard should give the comfort and feeling of a piano keyboard while playing music **(Ideal Final Result)**.

We need a piano-style keyboard to play music on computer and a conventional keyboard to type the text. We can have two different keyboards but that would take more desk space. Ideally the same keyboard should change to a conventional keyboard while typing text and should change to a piano-type keyboard while playing music **(Contradiction)**.

The invention uses a conventional keyboard as a piano keyboard by mounting a piano type keyboard on top of the conventional keyboard **(Principle-7: Nested doll)**.

**2.10 Adjustable and detachable mouse pad support with keyboard (Patent 6390432)**

**Background problem**

Mouse is the most common device to interact with the GUI applications. The mouse is normally placed on a mouse pad which is not ergonomic as it requires repeated arm movement from the keyboard to the mouse. Besides, different users like to keep the mouse in different positions and angles, for example the left handed users might put the mouse at the left side of the keyboard. There is a need to make an adjustable mouse pad which suites to both the hands, and reduces the arm movements as well.

**Solution provided by the invention**

VanderHeide Invented a new keyboard having a mouse pad support (patent 6390432, assigned to Knape & Vogt Manufacturing Company, May 02). The keyboard has a connector arm adapted for mounting to the keyboard support to hold the mouse pad. The joint typically includes ball and socket portions, which allows the mouse pad to move to any direction. The keyboard can have more than one sockets to connect the arm at different places to hold the mouse pad.

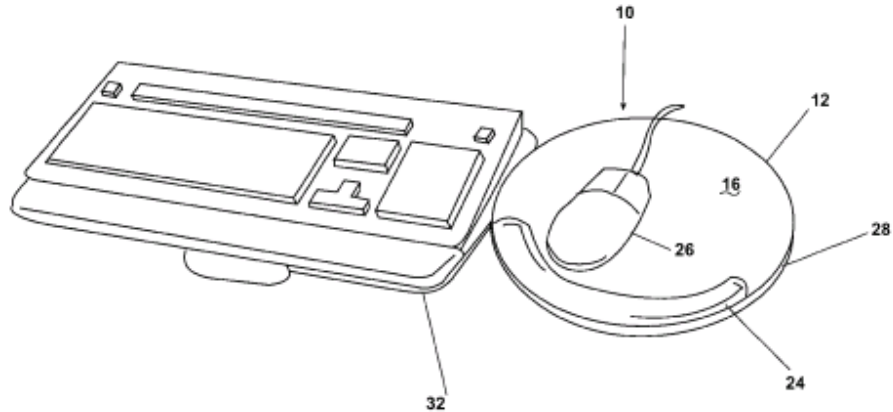

**TRIZ based solution**

The invention attaches a mouse pad support with the keyboard that reduces arm movement **(Principle-5: Merging)**.

The mouse pad is flexible to be attached to the left-hand side or right-hand side depending on the arm position **(Principle-15: Dynamize)**.

# 3. Summary and conclusion

Thus we can find that the keyboard is not only the keyboard but it can do many other things apart from the basic keyboard functions. When the keyboard is having an attachment, the basic functionalities of the keyboard are extended to achieve various other objectives. Some of the attachments we saw are;
- Keyboard template holding attachments
- Paper holder attachments
- Pointing device holder attachments
- Mouse pad attachments
- Wrist rest attachments
- CD ROM and other storage device attachments
- Multimedia attachments
- Joystick and other such device attachments, etc.

# Reference:

1. US Patent 4575591, Attaching a joystick with the keyboard thereby saving space. , Invented by Thomas Lugaresi, March 86.

2. 5096317, "Computer key cover apparatus", invented by Phillippe, issued Mar 1992.

3. 5281958, "Pointing device with adjustable clamp attachable to a keyboard", Ashmun et al., assigned to Microsoft, Jan 94

4. 5786861, "Clipboard attachments to computer keyboard", Invented by Joan Parker, July 98

5. 5881318, "Keyboard audio controls for integrated CD-ROM players", Invented by Liebenow, assigned to Gateway 2000, Inc., March 99

6. 5892502, "Keyboard having a storage device", Invented by Hiller Jeffrey, March 99

7. 5892503, "Multimedia console keyboard", Seung-Min Kim, assigned to AST Research, April 99

8. 5909210, "Keyboard-compatible optical determination of object's position", Knox et al, assigned to Compaq Computer Corporation, June 99

9. 5971635, "Piano-style keyboard attachment for computer keyboard", Herbert Wise, Assigned to Music Sales Corporation, Oct 99

10. 6390432, "Adjustable and detachable mouse pad support with Keyboard", VanderHeide, assigned to Knape & Vogt Manufacturing Company, May 02

11. US Patent and Trademark Office (USPTO) site, http://www.uspto.gov/